# Motifs of Networks from Frictional Interfaces

H.O.Ghaffari [1],R.P.Young

*Department of Civil Engineering and Lassonde Institute,*
*University of Toronto, Toronto, ON, Canada*

We have developed different network approaches to analyze complex patterns of frictional interfaces (contact area developments). Network theory is a fundamental tool for the modern understanding of complex systems in which, by a simple graph representation, the elementary units of a system become nodes, and their mutual interactions become links. With this transformation of a system into a network space, many properties of the system's structure and dynamics can be inferred. The rupture sequence of shear fractures were studied using a transformation form of contact patterns to complex networks; subsequently, sub-graph abundance within the corresponding networks was analyzed. To distinguish the different roles of collective deformation of an interface's elements, pure and non-pure contact patches (i.e., aperture) were mapped onto the nodes. The contact patches were connected with each other by using measurements of similarities as well as constrained geometrical distance and amount of net-contact area per patch, which yielded directed and non-directed networks. A universal trend in sub-graph distribution was observed. We confirmed that super-family phenomena are independent from rupture types in shear processes (as well as in slow or sub-Rayleigh fronts). Furthermore, global features of frictional interfaces as well as shear strength or hydraulic properties were scaled with motifs evolution. In particular, it was found that more common transitive motifs indicate residual shear strength stages, where fluctuations of stored potential energy surrounding rupture tip were minimal. Our approaches were tested over different available data sets, and it was found that discrete as well as real-time contact measurements resulted in the same universal patterns of sub-graphs.

[1] h.o.ghaffari@gmail.com; hamed.owladeghaffari@utoronto.ca
*Department of Civil Engineering and Lassonde Institute,*
*University of Toronto, Toronto, ON, Canada ; MB108-170 College Street, Toronto, Canada M5S3E3*
*Tel:+16475225025*

**Table of Contents**

## *Introduction*

Evolution of macroscopic friction in frictional interfaces originates from the sequence of contact area variations (Bowden and Tabor, 2001; Dieterich ,1994; Rubinstein et al, 2004; Ghaffari et al ,2010; Ben-David and Fineberg, 2010) . The formation and rupturing of new contact areas (i.e., bonds, junctions, asperities) forming between two surfaces results in stick-slip motion. Classical characteristics of stick-slip motion are fast frictional strength drops and energy released as spikes in ultrasonic wave forms (Ben-David and Fineberg, 2010). Recent findings also suggest that stick-slip motion is related to collective interactions of contact areas (Budakian and Putterman, 2000; Filippov et al, 2004; Rubinstein et al, 2009; Ghaffari et al 2009 and 2011) Hence, sheared systems-particularly in the form of granular elements-an anisotropic and long correlation around shear cracks (Maloney and Robbins,2009; Herrera et al ,2011; Chikkadi et al, 2011;Ghaffari et al 2011) .One way to characterize the contact patterns and their collective behavior is to "granulate" the patterns into separated "contact or non-contact" patches. The strings or patches, as the hypothetical objects, are the profiles constructed by collecting elements of the system (pure contacts or relative contact areas) along a thin line (i.e., a ribbon). Considering only pure contact cells represents stair-like patches, which we called contact patches or contact strings (Figure 1; the lower panel shows aperture patches). "Patches" are characterized by mapping the interactions onto networks, in which each patch is mapped onto a node and the corresponding interactions, connecting the nodes, are the links.

More recently, it was found that the characteristics of friction networks show a direct correlation with the mechanical (*i.e.*, shear strength) and hydraulic (*i.e.*, permeability) features of frictional interfaces (Also see supplementary documents figures S.2, S.3 and S.6). More specifically, close correlation of shear stress with the evolution of the formed modulus within the constructed networks have been proposed (Ghaffari et al, 2010; Ghaffari et al, 2011). From a different point of view, average shear stress indicates the stored potential energy surrounding rupture tip. Consequently, the flow of energy at contact zone (rupture zone) is related to cooperative interactions and the collective motion or deformation of the contacts or bonds. The storied energy in asperities is proportional with elastic properties of asperities and contact areas. The changes in contacts are expressed in terms of friction laws such as rate and state or Coulomb's friction laws (Dieterich, 1978). If we consider that there is a long-range correlation among particles ("elements") of the interface, then the patterns of energy propagation will be dramatically affected by correlation patterns. To better understand the nature of energy propagation, we note the importance of local structures in friction networks, determined by sub-graphs.

In this study we use sub-graph analysis of the pure contact patches or aperture patches (relative net contact areas) to show that the flow of energy (or entrapping) during a shear process is related to a super-family (universal) phenomenon, i.e., an identical trend in the distribution of sub-graphs. Two different approaches are used to construct patch networks. In the first approach, aperture patches are connected to each other by considering a similarity relationship. The second approach connects patches based on the similarity in number of contacts per patch and constrained geometrical distance. The networks resulting from the first method yield nondirected networks, while the networks resulting from the second approach are directed networks. Here, we exploit the evolution of the motifs of the nondirected networks as well as the directed graphs through shear slip. Until now, the ordering or direction of edges within contact patches has not been considered for constructing networks.

## *Materials and Methodology: Friction Networks*

To set up a nondirected network, we considered each patch of measured relative contact areas (i.e., aperture) as a node. Each aperture patch has *N* pixels, with each pixel showing the void size of the respective cell. The length of the profiles varies depending on the direction of patches. Our research takes into account two extremum directions, namely parallel and perpendicular patches related to shear direction. The maximum number of patches (in our cases) were observed in the direction perpendicular to the shear, while the minimum number is to be found in the parallel direction. To create an edge between two nodes, a correlation measurement ($C_{ij}$) over the aperture profiles was used. We assumed that if $C_{ij} \geq r_c$, then a link between two nodes was attached. Different approaches have been used toward this goal, including the observation link density, the dominant correlation among nodes, *C-K (clustering coefficient-node's degree);* or *P-Z (participation of edges in modules-internal connectivity of modules)* spaces and distribution of edges or clusters (Gao et al 2009 and 2010; Donges et al ,2009 ) . In the present study, we chose $r_c$ to be ~20 % of the maximum correlation value which nearly approximate the stable region of local flow of energy versus variation of threshold level (see supplementary document figure. S7 and figure S.8). With this threshold value, we considered relatively high correlated patches. It should be noted that applications of the correlation process in seismology as well as in the estimation of Green's functions, in modeling seismic coda, and in detecting new features of earthquakes have provided a powerful tool for data set analysis. From this perspective, characterization of such a correlation process (here conducted in the spatial dimension) with networks could be promising in terms of modeling complex geophysical phenomena. The choice of using aperture patches was inspired by the importance of aspect ratio in fracture length, i.e., a linear scaling of fracture growth with maximum aperture. From linear fracture mechanics, it can be proved that apertures show the index of available energy to rupture. Also, apertures can be assumed as relative contact areas where zero-apertures are pure contact areas (for example see Rubinstein et al, 2004; Sharifzadeh at al, 2008).

In the second approach, a directed network was constructed based on the capacity of patches to store energy *i.e.*, to entrap energy. The number of contacts per each patch indicates the preferentiality of energy flow, where energy prefers to be stored in the patches with maximum contacts. Based on the assumption of this preferentiality, a directed network over contact patches is set up (patches are constructed from aperture profiles -see figure 1d) with attributing the value 0 to aperture and the value 1 to contact cells). Further, to consider the spatial position of patches, a constrained geometrical distance was considered. The link is outgoing from patch *i* to patch *j* when $\left|C_{S_i}\right| < \left|C_{S_j}\right|$ and $|i-j| \le R$, in which $\left|C_{S_i}\right|$ stands for the number of contact cells in the $i^{\text{th}}$ patch and $R$ is the radius (nearest neighborhood) of the interaction. If $\left|C_{S_i}\right| = \left|C_{S_j}\right|$ and $|i-j| \le R$, there is no preferentiality in energy flow, and there will be a mutual edge (edges in both directions) at each node. It is noteworthy that with such a connectivity definition, we illuminate the entrapping of energy in absolute closed loops (for example, a three-node feedback loop – Fig. 4, index 7). Following this, the patterns of the energy diffusion around each patch are characterized by directed networks. As a complementary investigation, correlations between contact patches were separately substituted with the constrained geometrical distance. Our interest in this study is to capture the motif distribution of the directed networks; other characteristics of the directed networks will be reported elsewhere.

Sub-graphs are the nodes within the network with the special shape(s) of connectivity together. The relative abundance of sub-graphs has been shown to be an index of networks' information processing functionality. They also correlate with the global characteristics of the networks (Vazquez et al, 2004). The network motifs introduced by Milo et al. (Milo et al,2002; Milo et al.,2004; Alon,2007) are particular sub-graphs representing patterns of local interconnections between the nodes in the network. A motif is a sub-graph that appears more than a certain amount. It can be defined more precisely using Z-scores, which compare motif distribution with corresponding random networks (Milo et al., 2002; Milo et al., 2004). A motif of size *k* (containing *k* nodes) is called a *k*-motif (or generally sub-graph). In addition to applications of motif analysis over different networks in Milo et al, recently motif analysis on transformed forms of time series to networks revealed the method's capability to distinguish between and characterize different dynamic systems (Gao et al, 2010; Xu et al, 2008).

Throughout this study, based on the first approach described we used 4-point sub-graph distribution, while for the second networks, the triad profiles (3 point sub-graphs) are presented. We illustrate our approaches over 3 cases with two different normal stresses on the rock interfaces (Real-time contacts and their friction networks analysis can be followed in Figures S.12, S.13 and S.14 in supplementary document). The laboratory test procedure involved preparing a rough fracture, measuring the morphology of halves with a laser scanner, and measuring permeability. The rock used was granite with a unit weight of 25.9 $kN/m^3$ and a uniaxial compressive strength of 172 MPa. An artificial rock joint was made by splitting the specimen mid-height with a special joint-creating apparatus (Sharifzadeh, et al, 2005; Sharifzadeh et al 2008). The sides of the joint were cut down after it was created. The final sizes of the samples were 180 mm in length, 100 mm in width, and 80 mm in height. A virtual mesh with a square element size of 0.2 mm was spread on each surface, and the height at each position was measured with a laser scanner. The procedural details of reconstructing the aperture fields can be found in. The two aperture fields – corresponding with case 1 and case 2, respectively – were then sheared under 3 MPa normal stress (Normal force/Nominal area of interface). The first interface is a mated interface, in which the top and bottom surfaces were matched (Fig. 1a). Case 2 features an unmated interface, in which an initially significant non-contact zone in zero-slip was measured (Fig. 1b). Case 3 features a mated fracture with 5 MPa normal stress (Fig. 1c). All of the mentioned interfaces, under the specified normal loads, were sheared in parallel with the conventional X-direction, over 20 mm displacement.

## *Results and Discussion*

Interestingly, the resulting 4-point sub-graphs for parallel and perpendicular aperture profiles (figures 3 and 4) show roughly the same trend in motif rank distribution. In all of the distributions, a slight drop in index 2 and a fast drop in index 4 are observed. These two motifs are less ordered and connected, compared to other motifs. Motif 3 and 5 keep their dominancy in ranking over nearly all displacements, while index 6, when reaching the residual stages in shear stress (final displacements), assumes the same prevalence. In fact, the system evolves from a complete heterogonous rank to a less heterogonous one. The increment of motif 6 as the most transitive motif is more relevant in less complex energy flow patterns, in which the synchronization of information diffusion is much faster and congestion of contact areas is less pronounced (Xu et al ,2008; Gao et al, 2010; Lodato et al,2007). Over presentation of terrads 3, 5 can be related to possible fast transformation of energy through a cross-cut link. From this perspective, the obtained motif structures are very close to nondirected networks of protein structures with constrained geometrical connections of secondary-structure elements (Milo et al, 2004).

When plotting the evolution of individual motifs before transition to quasi-stable stages, general drops (or nearly no change) in motif ranks were observed. Particularly, this fluctuation occurs after shear stress drops (maximum Coulomb threshold). After this stage, a uniform growth in evolution of all motifs can be observed. The details of these drops differ between mated and unmated fractures, and the changes depend on the directions of the patches. For instance, a spike of the abundance of sub-graph patterns in SD=2mm to 3mm for case 1 is in agreement with the mechanical deformation and dilatancy of the fracture (see supplementary document for analysis of sub-graphs over real time contact measures) . Also, a drop in this case for all sub-graphs from SD=0mm to SD=1 mm is the index of a fast rate of energy storage in rupture tip where coincides with interlocking of asperities. The observed patterns indicate the interface's mechanical response (stored energy) as a global collection of patches, and its local topology, like sub-graphs, is interdependent (see Figures S.2, S.12 and S.13 in supplementary document). In other words, the large scale and global reaction of the interface to shear forces can be uncovered through the analysis of local subgraph structures. As a complementary inspection, the fluid flow patterns (for case 1) were modeled in parallel with the shear direction, and the permeability of the interfaces was compared with the evolution of the sub-graphs (Ghaffari et al, 2011). The results show that transitive motifs are biased to less complex flow, with minimum channelization and folding of flow patterns.

To construct directed networks, we created $R=10$ units (see supplementary document to discuss about $R$- figure. S9). Figure 4 shows the triad rank distribution over perpendicular profiles. As one can follow, a common feature in trend of sub-graphs frequencies observed for all of cases. Due to the special design of the networks, normally the motifs in feedback loops such as 7, 10 and 12 in three point sub-graphs and Bi-fan, Bi-parallel and four node feedback loops in four point sub-graphs were eliminated. In index 9, an increasing prevalence of mutual edges was encoded with the development of the shear forces and upon reaching the quasi-stable stage. However, the feed-forward loops (index 5) were dominant for all of the cases. This indicates that the configuration of contacts against shear forces occurs in a way that responds to the information signals- shear stress stimuli-is fast. We observed the same results for four-node motifs with a distinguished peak in double-feed forward loops. The second summit occurs in the three chain motif (index 2 or index 9), where it reflects the flexibility of networks against cascades, i.e., fast transferring of stimuli. From this perspective and the general trend of the triads, the obtained motif distributions are close to sub-graphs from signal-transduction interactions in mammalian cells and synaptic wiring of neurons (Milo et al, 2004; Alon, 2007). This feature, in addition to nondirected motif inspection and real time friction network analysis, shows a universal configuration of interfaces against shear forces, which form patterns in such a way as to allow information to transform and synchronize very fast. Sensitivity of the motif ranks was tested by changing the interaction radius $R$. For a broad range of $R$, the results showed that the trend of sub-graph ranks follow the same pattern (see supplementary document). Also, by using a correlation measure instead of a Euclidian metric over contact patches, we found a similar trend (with small differences in evolution of motifs) with a strong reciprocal scaling of individual motif evolution in the stored potential energy surrounding rupture tip (i.e., shear strength-Supplementary document figure. S10).

In summary, we found that the sub-graph ranks can be used as a tool to characterize the microstructures of shear ruptures. The local dynamic structures of contact patches as a result of development of shear stress were scaled with motifs evolution, indicating global mechanical scaling with local properties of contact patches. More common transitive motifs were the indication of shear strength residual stages, in which the fluctuations of stored potential energy surrounding rupture tip were minimal. The natural configuration of contact areas imposes fast transformation of information through contact patches in which the feed-forward loops and successive chains of motifs control the functionality of energy flow.

❖ We would like to acknowledge and thank Prof. M. Sharif (Tehran Polytechnic), Prof. J. Fineberg (The Racah Institute of Physics, Hebrew University of Jerusalem) and S. Maegawa (Graduate School of Environment and Information Sciences, Yokohama National University, Japan) who provided the data set employed in this study.

*The supplementary document provides the results of the network analysis and motif distribution for real time contact measure.*

# *Figures*

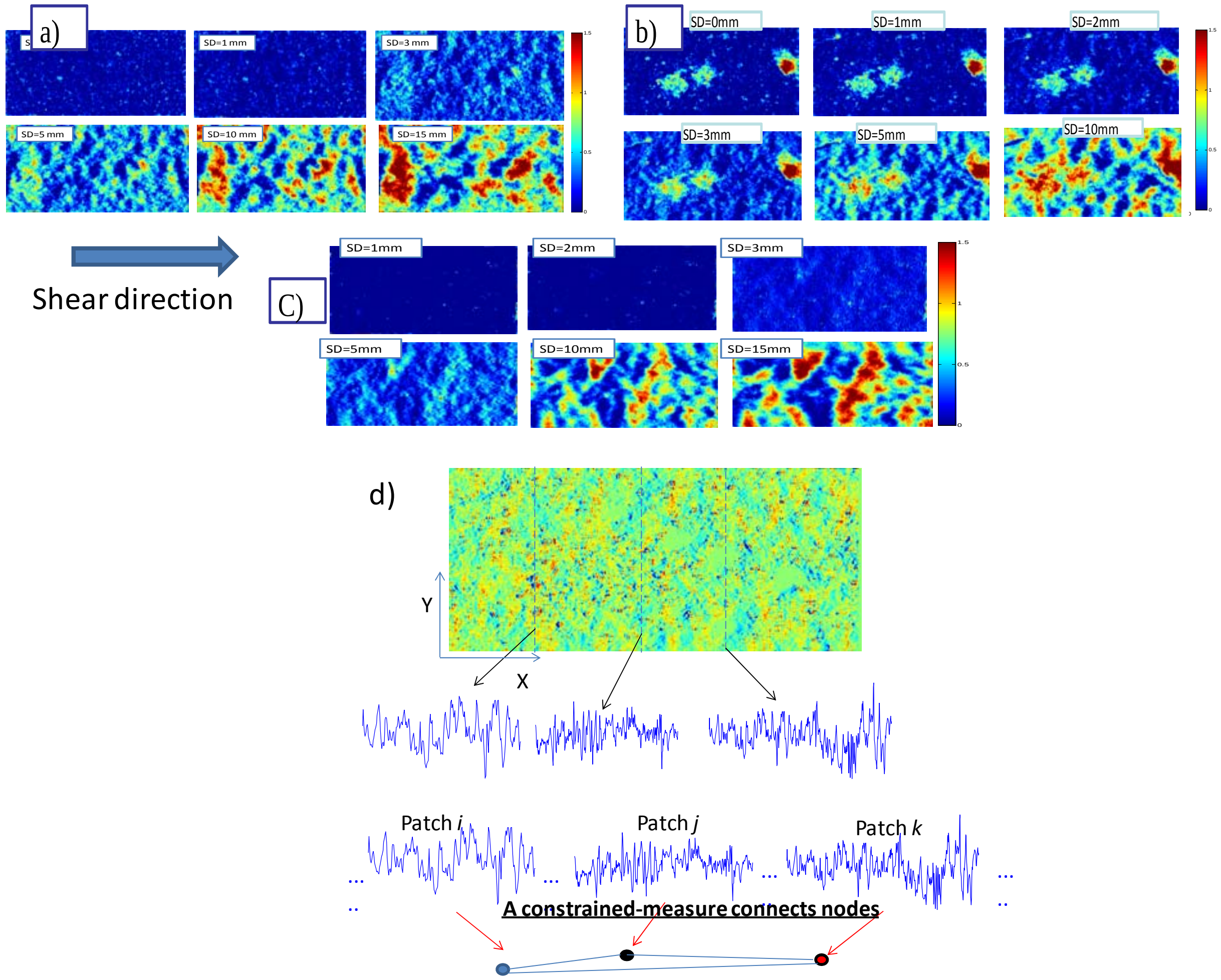

Figure 1. Measured aperture distribution within rock interfaces with different normal stress through 20 mm shear displacements (SD): a) Normal stress 3 MPa (case 1) ; b) Normal stress 3 MPa (case 2); and c) Normal Stress 5 MPa. The shear direction is from the left to the right side; d) Mapping a surface into the network space.

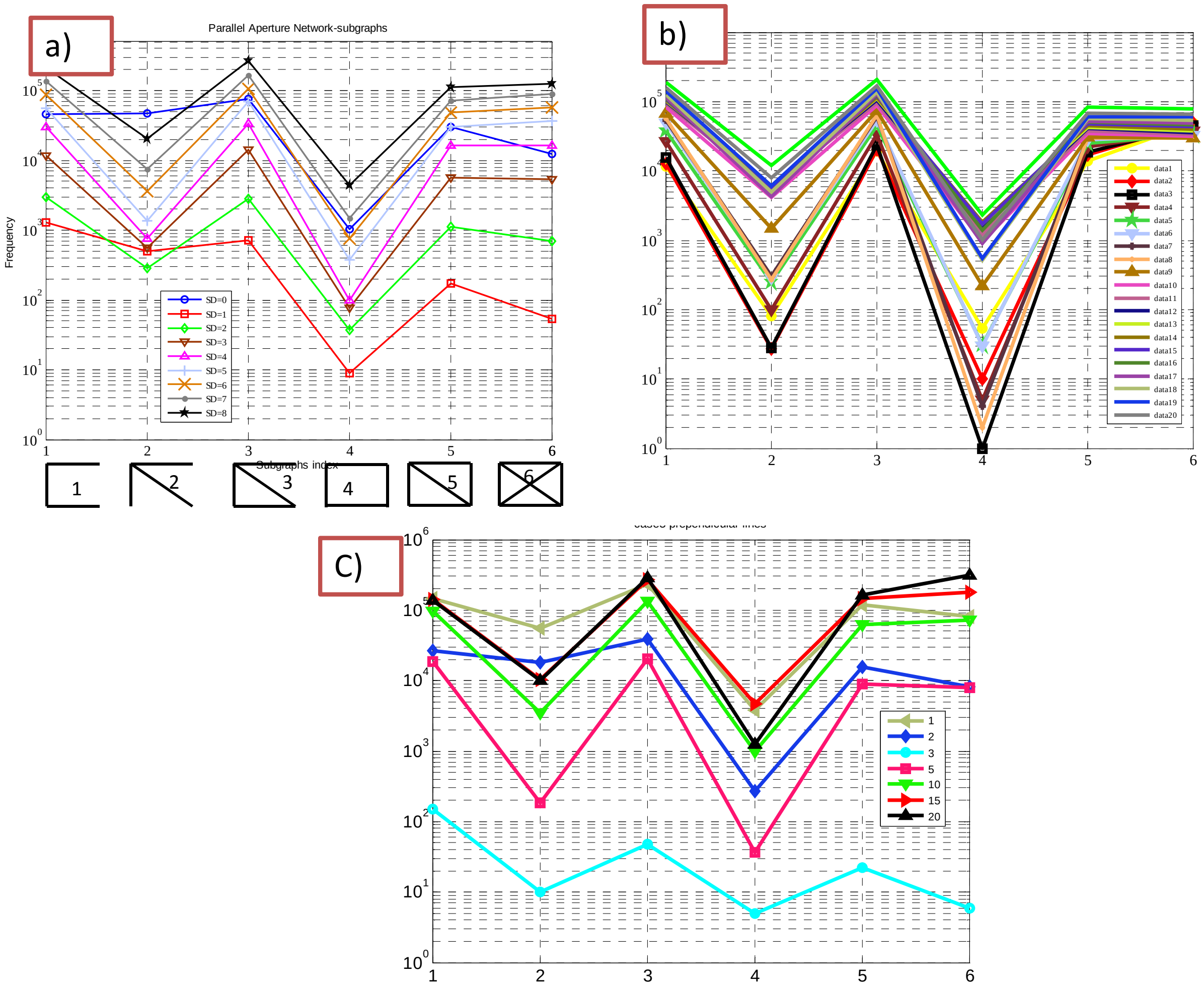

Figure 2. Distribution of 4-point sub-graphs over *nondirected networks* and three cases (see Fig.1): a) case 1; b) case 2; and c) case 3. Aperture patches are in *parallel* with the shear direction. Vertical axis shows frequencies of sub-graphs, and the X-axis is corresponding to the motif's indexes.

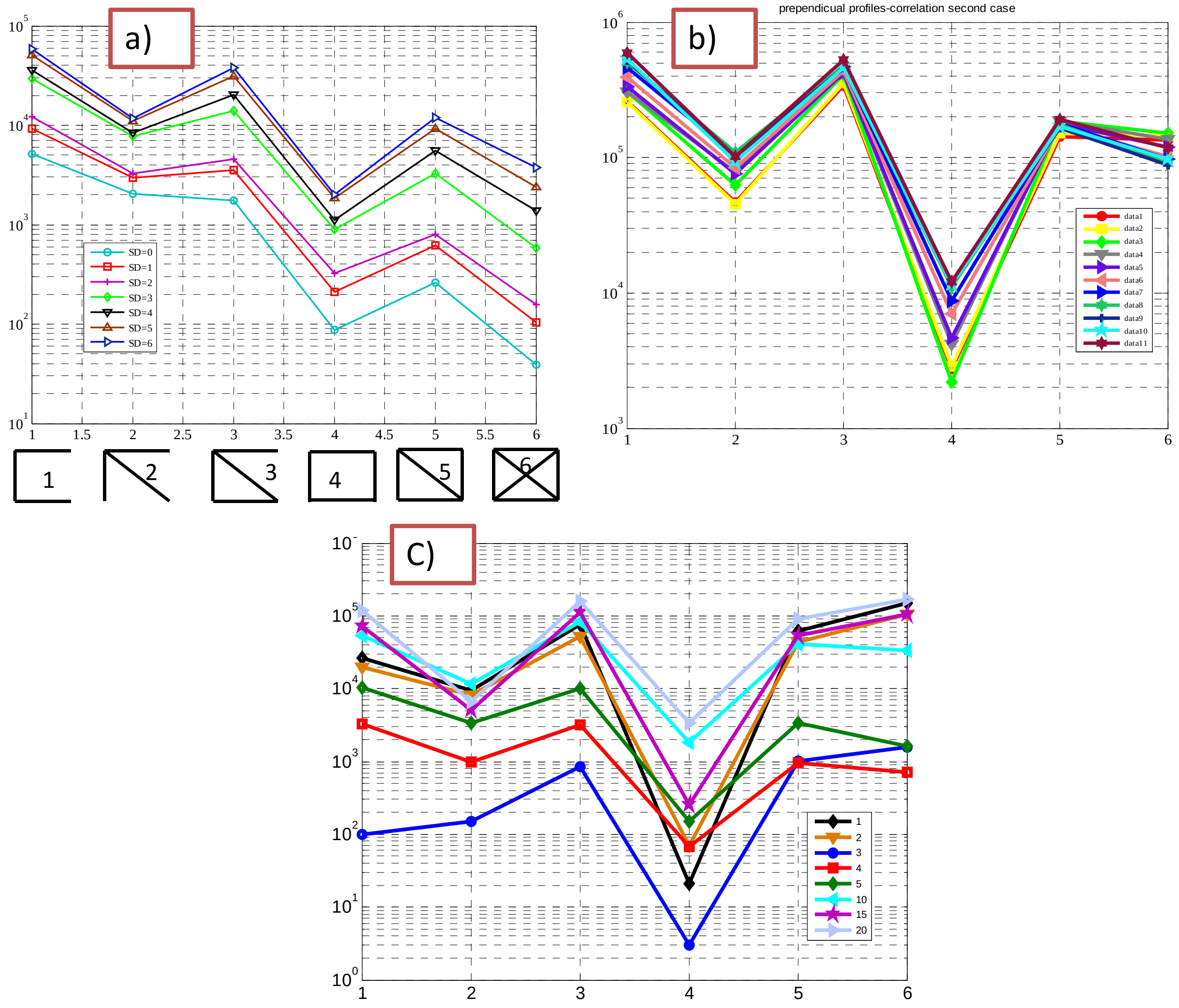

Figure 3. Distribution of 4-point sub-graphs over *nondirected networks* and three cases (see Fig.1): a) case 1; b) case 2; and c) case 3. Aperture patches are *perpendicular* with the shear direction.

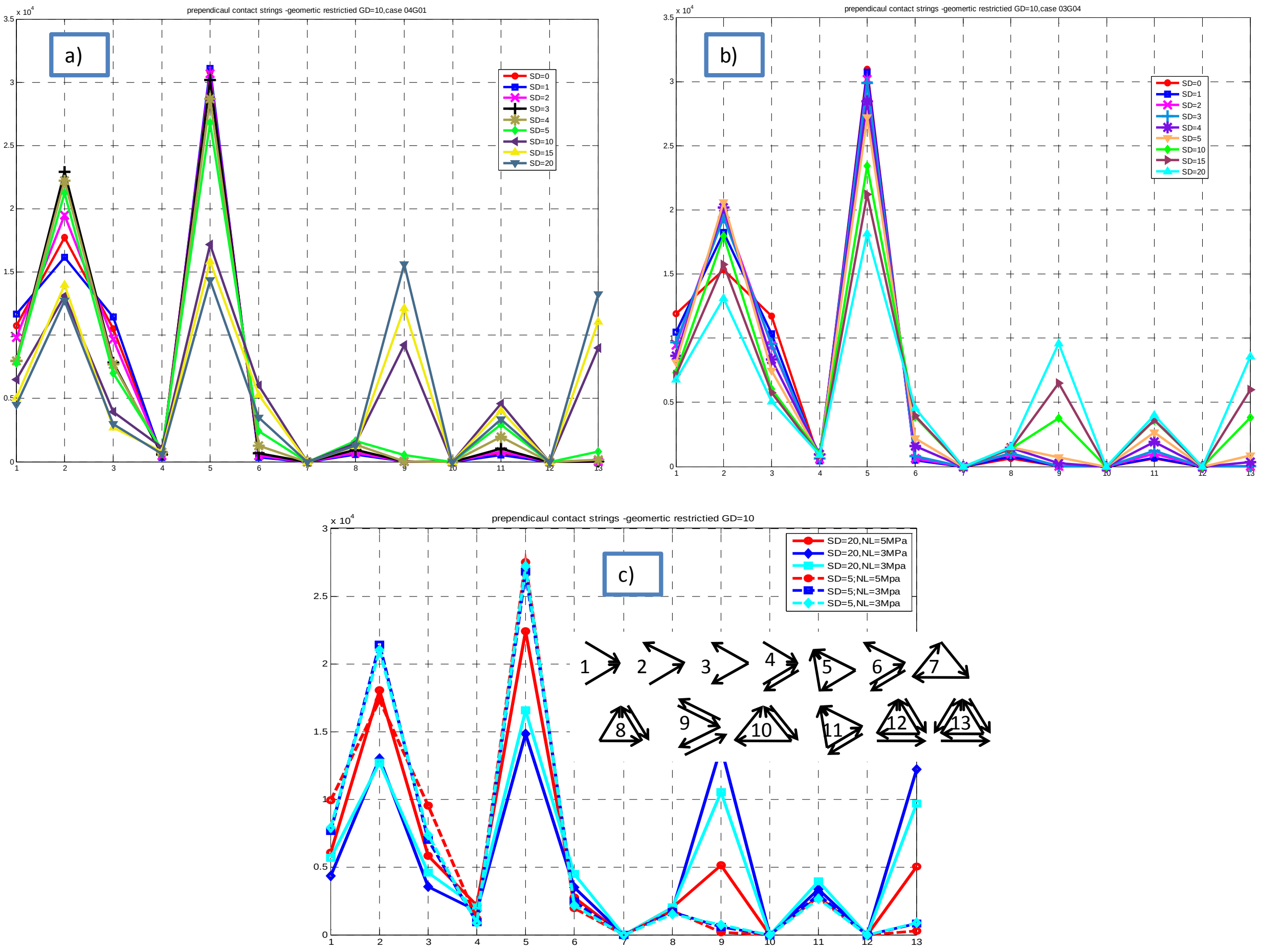

Figure 4. Distribution of 3-point sub-graphs over *directed–perpendicular networks*: a) case 1; b) case 2; and c) comparison of sub-graphs ranks in two slip displacements (5 and 20 mm) through 3 cases.

**Supplementary Document:**

# Motifs of Networks from Frictional Interfaces

H.O.Ghaffari [1],R.P.Young

*Department of Civil Engineering and Lassonde Institute, University of Toronto, Toronto, ON, Canada*

In this supplementary document, we describe more properties of the demonstrated friction networks. The first part covers more results from aperture patterns analysis and their relevance in friction analysis. We also, discuss about the threshold levels –employed in the main text to extract the sub-graphs. The second part includes the analysis of real-time contact patterns.

## Contents

[1] h.o.ghaffari@gmail.com; hamed.owladeghaffari@utoronto.ca
*Department of Civil Engineering and Lassonde Institute, University of Toronto, Toronto, ON, Canada ; MB108-170 College Street, Toronto, Canada M5S3E3*

*Tel:+16475225025*

# Friction Networks & Aperture patterns

## General Characteristics of Complex Networks

To proceed, we use several characteristics of networks. Each node is characterized by its degree $k_i$ and the clustering coefficient. Clustering coefficient (as a fraction of triangles (3 point loops/cycles) is $C_i$ defined as $C_i = \frac{2T_i}{k_i(k_i - 1)}$ where $T_i$ is the number of links among the neighbours of node *i*. Then, a node with *k* links participates on *T(k)* triangles. Furthermore, based on the role of a node in the modules of network, each node is assigned to its within-module degree (*Z*) and its participation coefficient (*P*). High values of *Z* indicate how well-connected node to other nodes in the same module and *P* is a measure to well-distribution of links of the node among different modules (modules can be assumed as the clusters .groups or communities in networks). To determine modularity and partition of the nodes into modules, the modularity *M* (i.e., objective function) is defined as [1]:

$$M = \sum_{s=1}^{N_M} [\frac{l_s}{L} - \left(\frac{d_s}{2L}\right)^2], \qquad (1)$$

in which $N_M$ is the number of modules (clusters) , $L = \frac{1}{2}\sum_i k_i$ , $l_s$ is the number of links in module and $d_s = \sum_i k_i^s$ (the sum of nodes degrees in module *s*) . With using an optimization algorithm, the cluster with maximum modularity is detected.

## The Necessity of the Study

We discuss about the necessity of study of contact patterns (or aperture, asperity patterns). We present a simple mathematical analysis with respect to contact mechanics and asperities interactions (for details see [Johnson, K.L., Contact Mechanics. (Cambridge University Press, Cambridge, 1987) and doi: 10.1038/nature10589]. Let us assume two cases where two similar circular-contact areas (with radius *a*) formed between a rigid top surface and a linear half space with shear Modulus *G* and Poisson's ratio $\nu$ (~relative contact areas are same; this is equal somehow to observed aperture). The mechanical interactions between two asperities are approximated by Cerruti solution (we assume the contacts are under lateral shear force):

$$\begin{bmatrix} A^1 & 0 & B_x^1 & C^1 \\ 0 & A^1 & C^1 & B_y^1 \\ B_x^1 & C^1 & A^1 & 0 \\ C^1 & B_y^1 & 0 & A^1 \end{bmatrix} \begin{bmatrix} F_{x1}^1 \\ F_{y1}^1 \\ F_{x2}^1 \\ F_{y2}^1 \end{bmatrix} = \begin{bmatrix} \delta_{x1}^1 \\ \delta_{y1}^1 \\ \delta_{x2}^1 \\ \delta_{y2}^1 \end{bmatrix} \rightarrow \mathbf{MF} = \boldsymbol{\delta} \tag{2}$$

in which,

$$\begin{aligned} & r_1 = \sqrt{(x_2^1 - x_1^1)^2 + (y_2^1 - y_1^1)^2}, A = \frac{1}{8a}\frac{2-\nu}{G}, \\ & B_x^1 = \frac{1}{4\pi G}((2-2\nu)\frac{1}{r_1} + 2\nu\frac{(x_2^1 - x_1^1)^2}{r_1^3} \\ & B_y^1 = \frac{1}{4\pi G}((2-2\nu)\frac{1}{r_1} + 2\nu\frac{(y_2^1 - y_1^1)^2}{r_1^3} \\ & C^1 = \frac{1}{4\pi G}(2\nu\frac{(y_2^1 - y_1^1)(x_2^1 - x_1^1)}{r_1^3} \end{aligned} \tag{3}$$

The second case is similar to the first case (see Figure 1 b), however, a small difference in the position of the second asperity is induced. Further assumption in $x_1^1 = x_1^2 = y_1^1 \equiv 0; x_2^1 = x_2^2 \equiv 0$ where subscript and superscripts are local position index and case number (1 or 2), respectively. For a certain displacement and both case, shear force distribution among contact areas can be solved. If we assume, furthermore, $y_2^1 < y_2^2 \rightarrow r_2 > r_1$ then $A_1 = A_2; r_2 > r_1 \rightarrow B_x^1 > B_x^2, B_y^1 > B_y^2$. The later inequality implies the final shear stress distribution is different for each case as if we only change the pattern of contacts. For complex contact patterns where we have different contacts areas and connectivity of zones, **M** is so complex and analytically cannot be inferred. Then, study of friction is in a direct relationship with study of contact patterns rather than pure magnitude of contact areas. With a simple transformation of the surface into a network (graph) frame, we see the connectivity patterns of the network are dramatically change as well as other characteristics of the networks. In addition to this, we distinguish the observations of long-range correlations in sheared –granular systems which have been approved numerically and experimentally [2, 3]. Long-range correlations in the fluctuations of shear strain induce the "excitation of additional elastic modes". The strong correlation in the direction of shear lowers the effective resistance to rupture in the direction of shear. We characterized this correlation with networks.

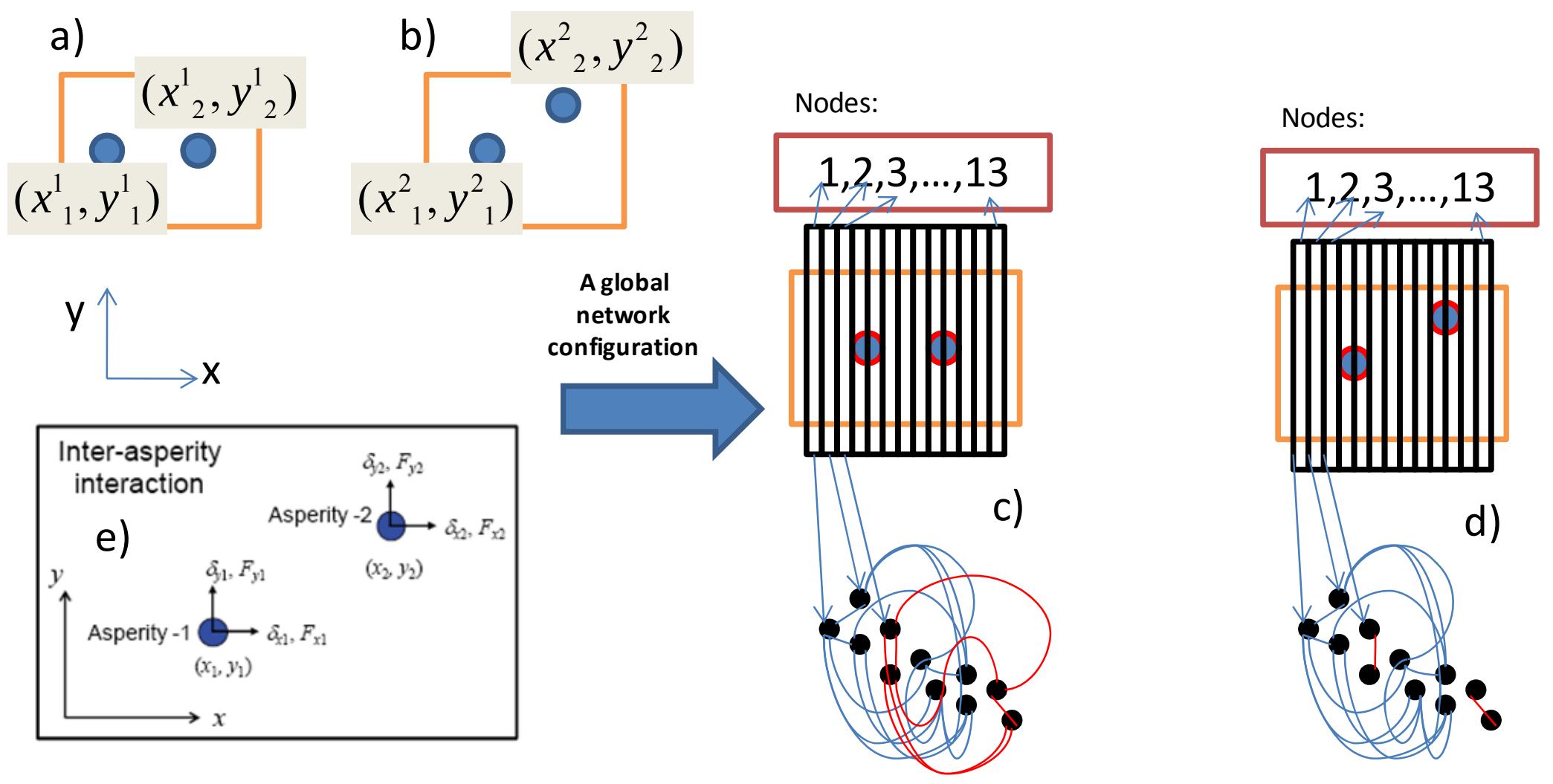

Figure S1. a,b) two similar circular contact areas with radius rand with different distance set up in a rigid plate.; c,d) transforming profiles (black rectangulars) to nodes and connecting them based on similarity of profiles(~ribbons) ;e) the components of forces and displacements due to mechanical interaction among contact areas.

## Some Characteristics of Friction-Aperture Networks

In Figure 2, we have shown the evolution of aperture patterns and also the developments of the related correlation patterns. In this part we focus on the highlighted patterns and obviously our results can be investigated to other patterns (presented patterns in Figure 1-main text). Obviously, one can follow a self-organized correlation patterns which can be analysed in terms of nonlinear dynamics methods. One of the methods to describe the self-organization nature of patterns is using a series of coupled equations in terms of inhibitor-activator (Turing patterns). Then, we can describe the evolution of number of edges (or node's degree) with respect to Turing patterns, which it reads:

$$\begin{aligned}
&\frac{\partial}{\partial t}k_i(t) = f(k_i,u_i) + \varepsilon_k \nabla^2 k(x_i,t) \\
&\frac{\partial}{\partial t}u_i(t) = g(k_i,u_i) + \varepsilon_u \nabla^2 u(x_i,t) \\
&f(k_i,u_i) = \gamma(a - k_i + k_i^2 u_i) \\
&g(k_i,u_i) = \gamma(b - k_i^2 u_i)
\end{aligned} \qquad (4)$$

in which $k_i$, $u_i$, $\varepsilon_k$ and $\varepsilon_u$ are node's degree ,a system attribute (as well as displacement or temperature) mobility of patterns coefficients ,respectively. With $\varepsilon_k$=1 and $\varepsilon_u$>1, we obtain

Turing patterns (a document describing possible real Turing patterns induced by friction is described in: Langmuir. 2011 Apr 19; 27(8):4772-9. Epub 2011 Mar 29).

A consideration of 3-point cycles (triangle loops) versus the nodes' degree shows a power low scaling (Fig. 3a):

$$T(k) \sim k^{\beta}, \qquad (5)$$

where the best fit for the collapsed data set reads $\beta \approx 2 \pm .3$ (which we call a coupling coefficient of local and global structures). With some mathematical analysis, one can show that adding $m$ edges increases the number of loops with $\beta^2 m^{\beta}$, which indicates a very congested structure of global and local sub-graphs during shear rupture. Also, we notice $C(k) \sim 2k^{\beta-2}$, so that for $\beta < 2$, a possible hierarchical structure can be predicted (see figure 3e). It is not worthy; we observe the classical hierarchical structures in friction networks in residual stage of friction (figure 3.b). The onset of the sliding is correlated with the drop of power law and Stick (interlocking) period characterized with sudden drop of $\beta$. Betweenness centrality (B.C) is a measure of how many shortest paths cross through a node and can be assumed as a local information flow in networks. The Scaling of mean values of B.C with clustering coefficient is remarkable and can be use to further friction laws in terms of networks parameters (figure 3.g). Increasing clustering coefficient values-while occurs in the residual stage- lowers local energy flow, indicates highly correlated apertures patches induces less local energy dispersion. In [4], we have also extensively described the evolution of super-nodes (nodes with high edges or hubs) and their roles in friction networks. For instance, in figure 3 h, the assortative mixing parameter has been shown. Assortativity is an index which shows how hubs can absorb other rich nodes. As one can follow first of all, the obtained friction networks are assortative and It means hubs attach to hubs; however transition from slip stage (stick part) to sliding step follows with decreasing assortativity and reaching to residual part occurs after 6 mm shear slip. We notice increasing apertures- in residual part-dramatically increases assortativity.

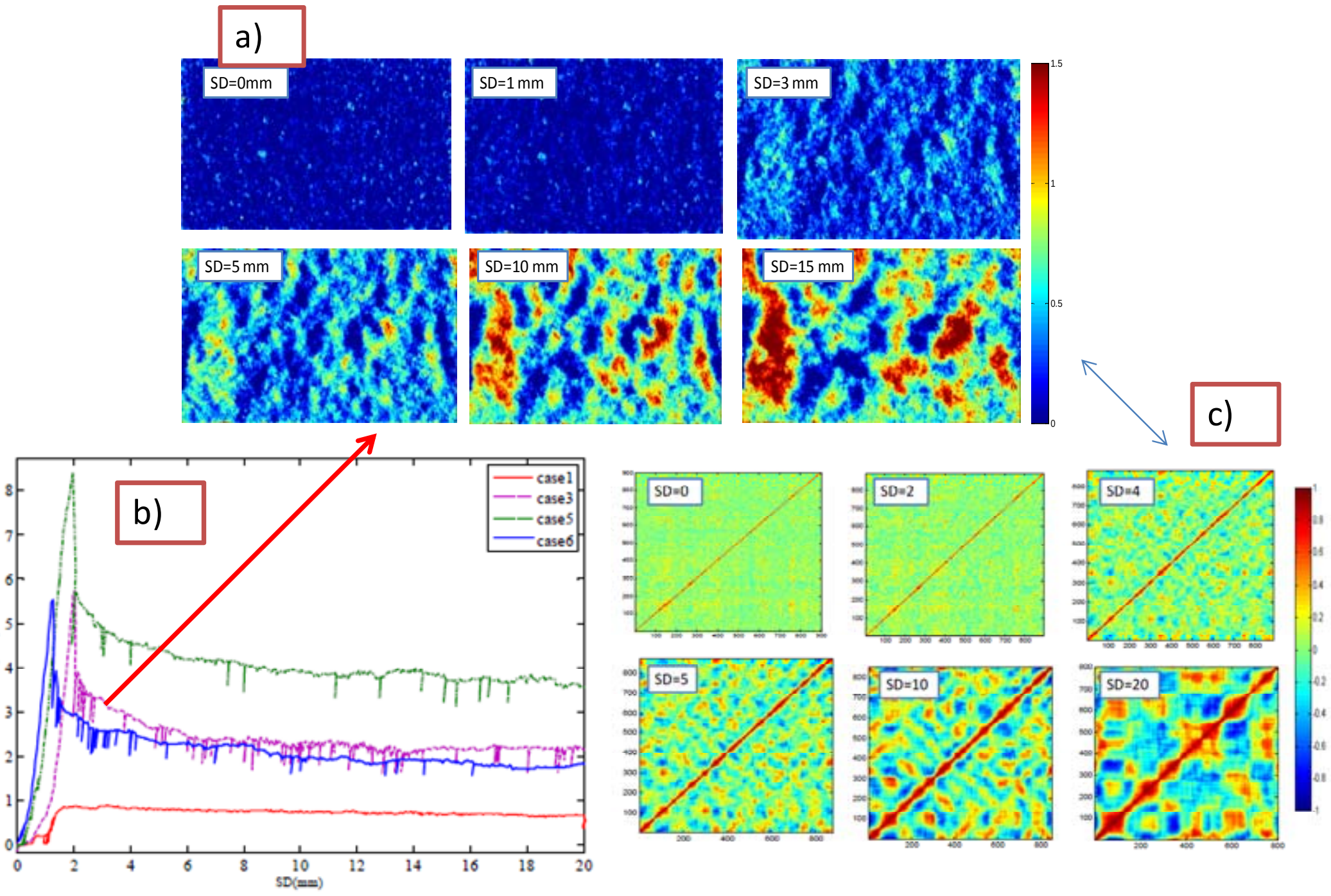

Figure S2. a)aperture patterns of a rock interface under 3Mpa normal load ;b) shear strength developments versus shear displacements (SD in mm);c) correlation patterns of aperture patches in perpendicular aperture profiles.

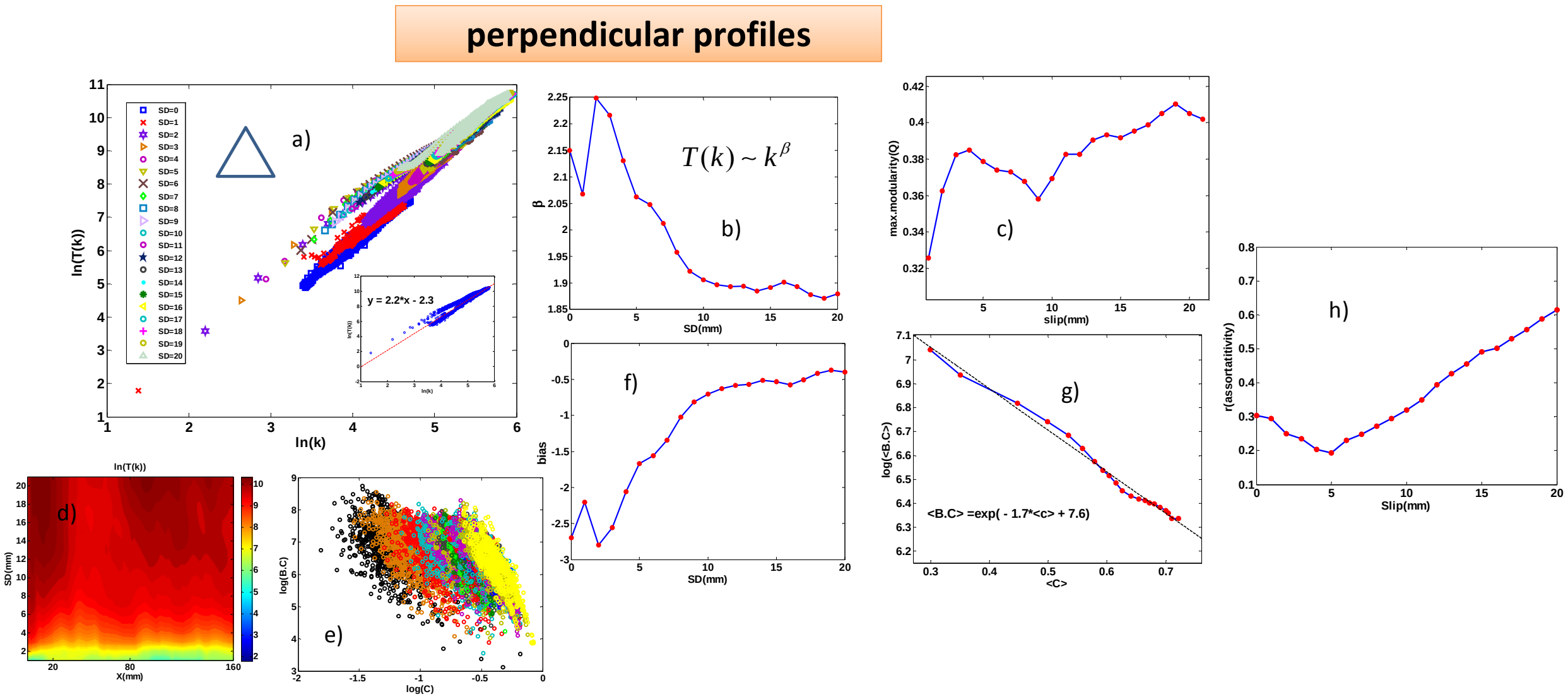

Figure S3.a) scaling of motif (3 point sub-graph) with number of edges for perpendicular friction-networks; b) details of power in scaling law: the onset of sliding is correlated with drop of power law. Stick (interlocking) period characterized with sudden drop of $\beta$ ; decreasing $\beta$ increases the bias parameter in exact scaling (linear in log-log coordinate) relation (see (f)). (c) Modularity of friction networks (maximized Q);(d) logarithm scale of loops variation with slip and along X;(e) scaling of betweennness centrality (as a local measure of information flow) with clustering coefficient ;(f) variation of the bias parameter in exact scaling(linear in log-log coordinate) relation (g) scaling of mean betweennness centrality (as a local measure of information flow) with mean clustering coefficient over all nodes (aperture patches) shows an exponential scaling law for all shear displacements with clustering coefficient indicates increasing clustering coefficient decreases local information flow; (h) assortativity of networks

with slip; It shows generally our networks are assortatitve as if releasing entrapped energy in contact areas lowers(decreases) assorattivity.

## Roughness and Friction Networks

Characterization of aperture of a rock joint has been realized by using several techniques such as employing statistical methods that include root-mean square (RMS), RMS of first derivative (Z2), RMS of second derivative (Z3), and structure function (SF) [5] ; geo-statistical methods to study spatial variation of asperity heights as well as spatial correlation (semi-variograms) [5] and ; correlation length which gives an idea of the aperture (dramatically) changes over the rock joint surface; the fractal models mainly used for scale effect analysis [5]. Obviously, the distribution of the network characteristics, indeed, can be used as another statistics to capture the backbone of friction patterns. In this section, we present the possible relation among two characteristics of the networks (clustering coefficient and node's degree) with roughness parameter.

We use RMS (root mean square) of profiles in two directions (parallel and perpendicular to shear). For the simplicity, we consider one surface (rupture surface) as a reference to calculate roughness value. In figures 4 and 5, we have illustrated the values of the RMS are plotted versus *k* and *c*. In initial time steps (shear displacements) the rupture surfaces are much smoother rather than the last shear slips and then for small variation of roughness we observe a significant change in the network characteristics. This is remarkable, because with RMS or fractal dimension (characteristic length) we may cover a part of complexity in the patterns of the frictional interfaces. However different characteristics of networks reveal different aspects of complexity and correlation of aperture or contact patches (corporative motion of particles). Hence, perspective as if roughness value gives an index to general fluctuation of surface heights (which somehow universal for tensile fractures; see [6]), however cannot say anything about possible correlated elements role in the evolution of sheared interfaces. Interestingly, we can cover characteristic length of the rupture surfaces with network parameters, such as modularity parameters. In other words, with using self-similarity concept over rupture surfaces the characteristic length can be defined as the critical length beyond which geometric irregularity of the shear fracture surface no longer exhibits the self-similarity (related to fractal dimension). To show how the characteristic length can be defined by networks parameters in Figure 6, we have shown the results of our calculations over two different type of the data set (for details of real time contact measure see section 2). The periodic patterns of the networks parameters for both case studies (aperture patches and relative contact's areas) indicate a dominant self-similarity in rupture surface. This can be as another improvement of our formulation for friction networks where the characteristic length as an index of asperities size is crucial and essential parameter.

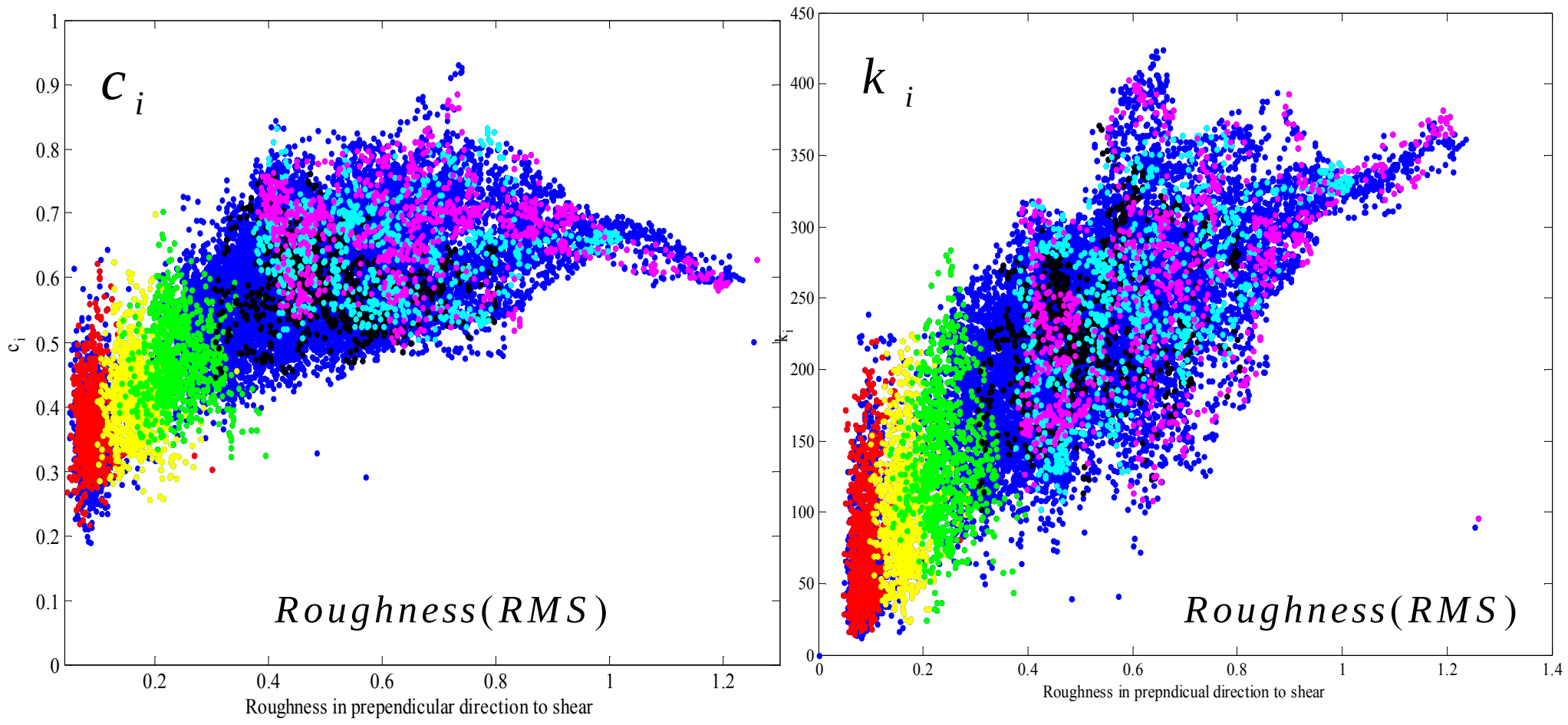

Figure S4. Roughness variation versus clustering coefficient (C) and node's degree (k) for perpendicular aperture profiles. The coloured points show the shear displacements (blue for last slip and red for initial displacement).

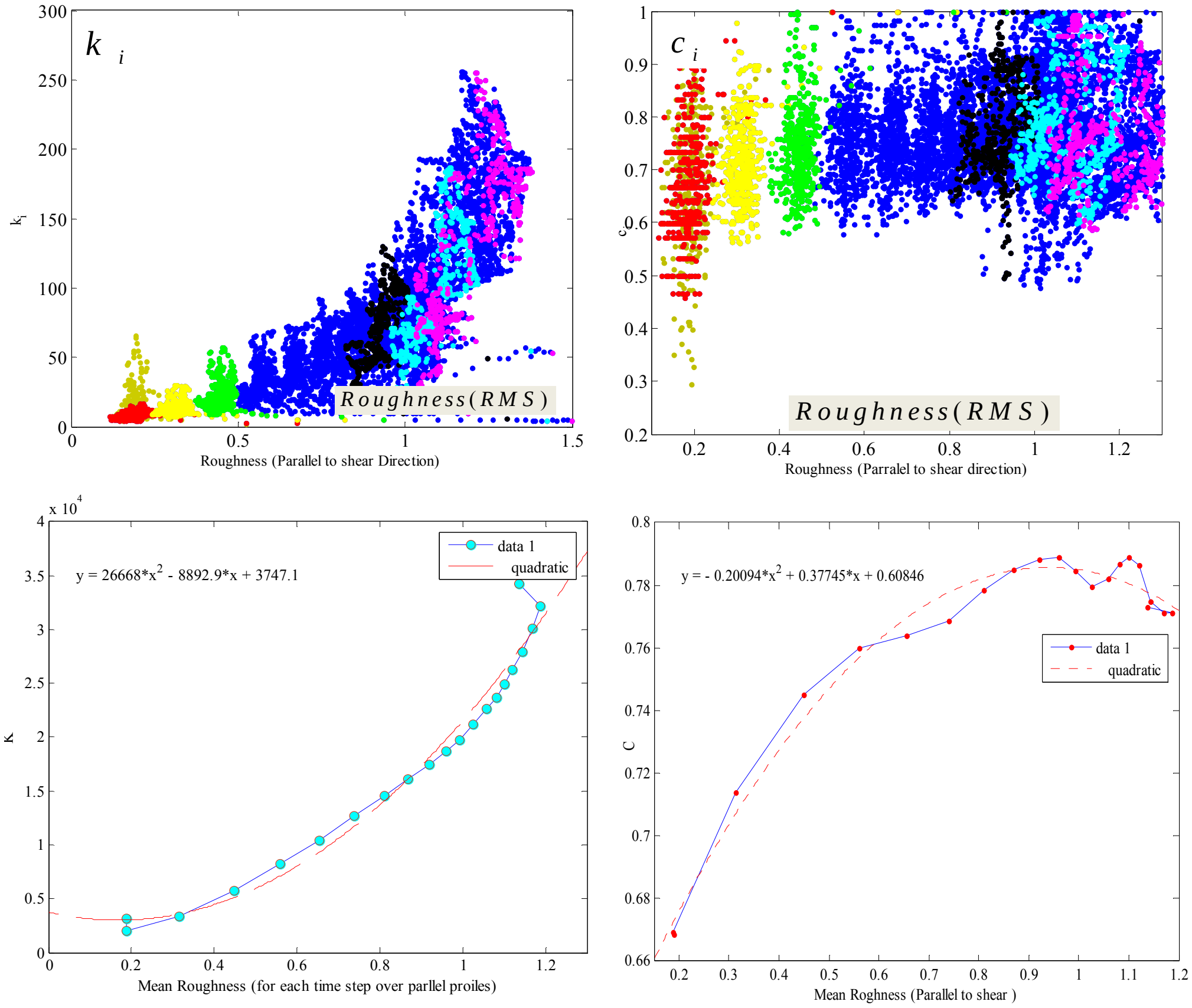

Figure S5. Roughness variation versus clustering coefficient ($c$) and node's degree ($k$) for parallel aperture profiles. The coloured points shows the shear displacements (blue for last slip and red for initial displacement). In down panel, we fitted the best polynomial function among mean RMS and <k> and <c>. This indicates generally increasing RMS index of roughness is compatible with increasing global characteristics of the networks.

**Periodic patterns; characteristic length of rupture surfaces**

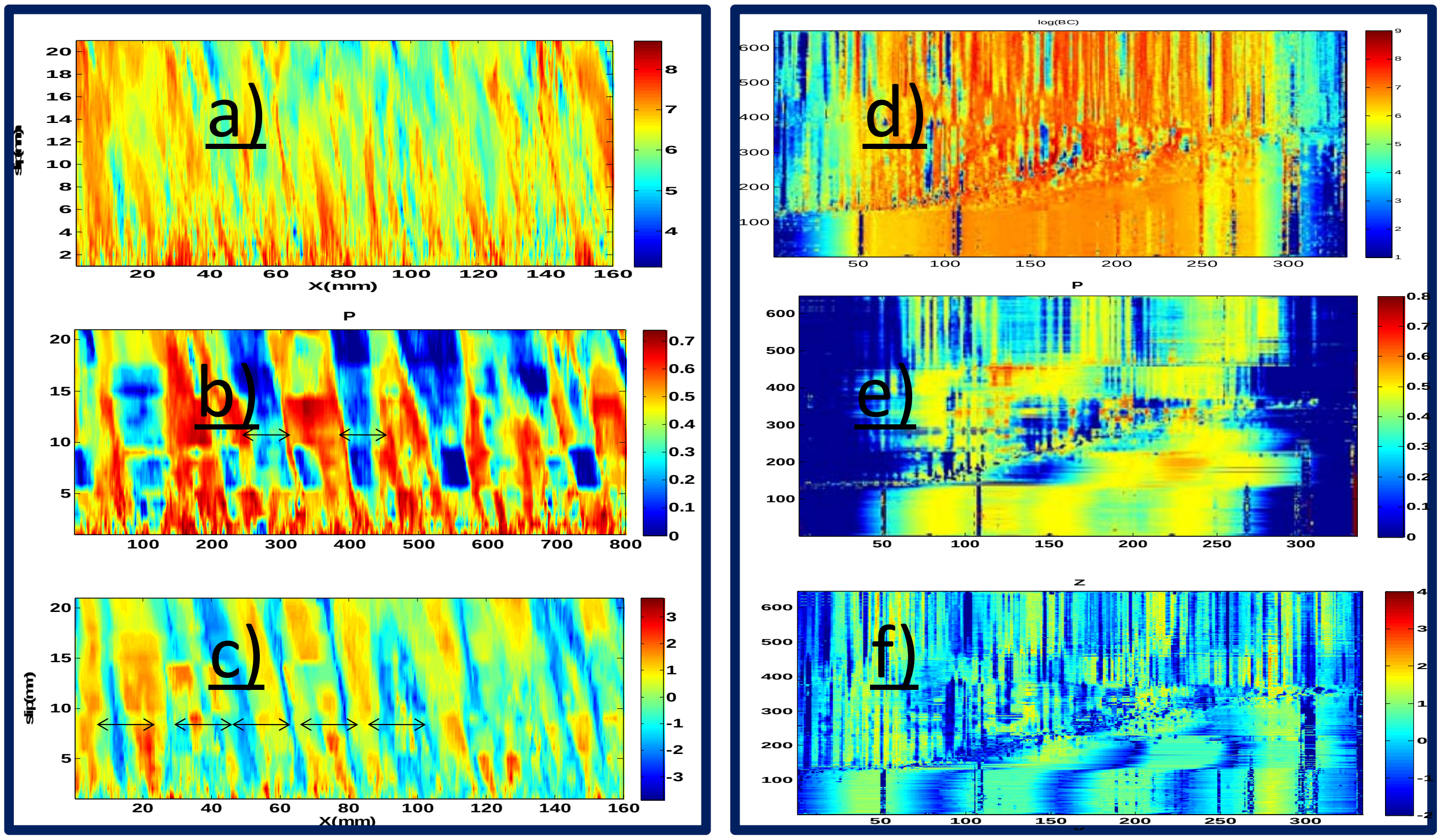

Figure S6. Periodic patterns as the characteristic length of rupture surface: a) logarithmic scale of betweeness centrality (B.C) through perpendicular aperture profiles(the vertical axis is shear slip) ;b) participation coefficient (P) in slip-X space ;c) within module degree (Z) in slip-X space ;d) ) logarithmic scale of betweeness centrality (B.C) through perpendicular aperture profiles(the vertical axis is time in pixels unit and X axis is 140 mm-see [7]) in real time contact measure which shows the onset of the motion ;b) participation coefficient (P) in slip-X space ;c) within module degree (Z) in slip-X space. We estimate the characteristic length for aperture profiles as wavelength 1-2 mm.

## Discussion : Threshold Levels

Selecting the threshold value is a critical value in our algorithm to extract friction networks. One the well-known methods in selection of threshold level is in association with networks of time-series and recurrent series analysis (i.e., density of edges). In figure 7, we have shown the sensitivity of the density of edges and betweenness centrality to $r_c$. To choose $r_c$, we used a nearly stable region in the betweenness centrality (B.C) - $r_c$ space (Fig. 7d), which is in analogy with the minimum value in the rate of edge density (Fig. 7c). Betweenness centrality is a measure of how many shortest paths cross through a node. This method has been used successfully in analysis of time-series patterns in network spaces. This interval is nearly equal to a choice of $r_c$ to be about 10-20 % of the maximum correlation value. The mentioned aperture patches may be observed in a manifold space. It turns out around the neighbourhood of $\frac{dk}{dr_c} \to 0$ the variation of B.C with $r_c$ is nearly constant (see figure 8). In other words, with choosing constant proper threshold level, we can reach to stable local information flow (consider our approach consider all slip displacements). The obtained approximate interval reveals the best threshold values where we can reach to backbone of the characteristics of friction networks. Recent studies also use different spaces as well as motif characteristics (normalized maximum size of sub-graphs). In figure 8e-f, I have presented two spaces variations with the truncation value (in shear slip 10mm). Obviously, the lowest threshold level (here -0.6), yields a congested and dense network where we only observe critical abnormalities in un-correlated patterns. With increasing threshold level, dispersion and density of the points increase so that for instance the general trend of points in *ln(B.C)-k* or *ln(B.C)-c* space is nearly the same. We notice for different shear slips around $r_c$ = -0.1-0.2, the local information flow is getting a nearly stationary level (also see figure 8 g), and then choosing a constant threshold level does satisfy our goal (reaching to stable structure (non-random)). Obviously, one can fully characterize the mentioned method and compare with edges density method. It is noteworthy with increasing threshold level, we gradually loose the patterns of correlation where only the profiles around each profiles is considered (influence radius/see diagonal components of the correlation matrix in figure 2c). The latter critical value filters all relative correlated elements and only picks very high correlated profiles.

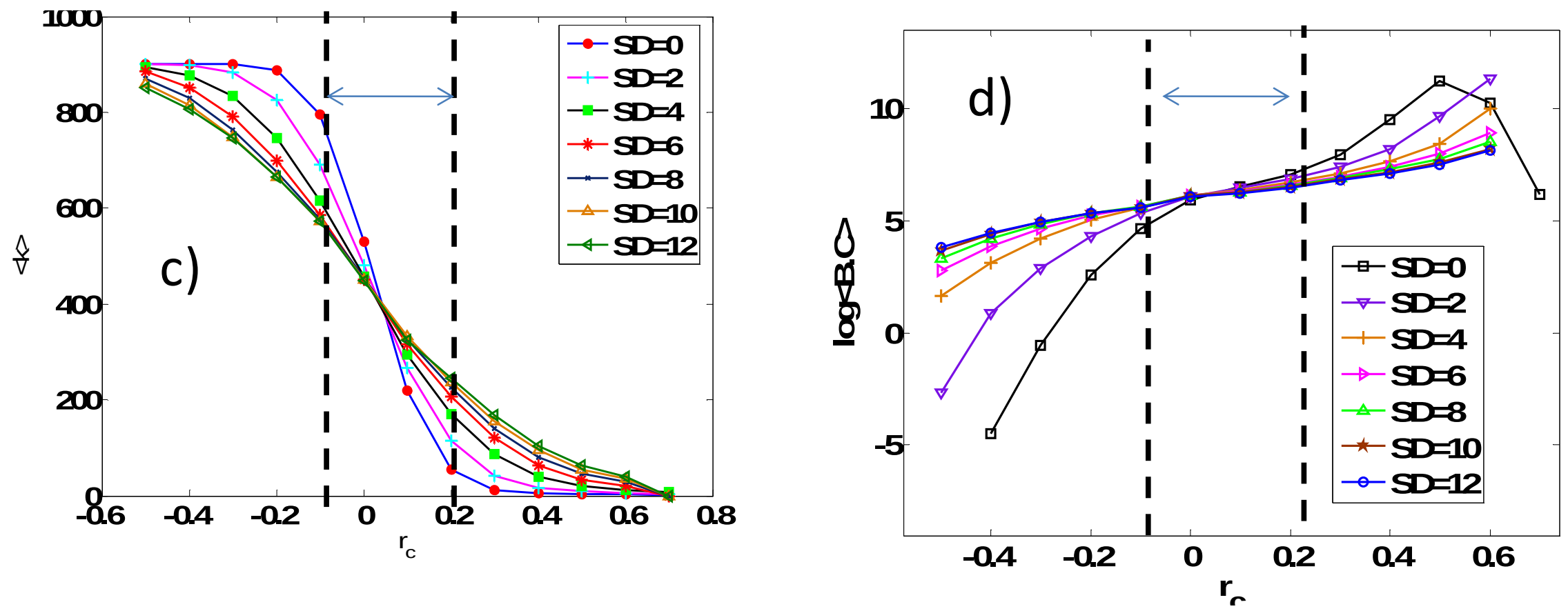

Figure S7. c) density of edges versus threshold level and d) natural logarithmic variation of mean betweennness centrality (B.C) with truncation value. The indicated interval with arrows shows the best possible threshold level where the minimum variation of log<*B.C*> occurs (the most stable-dominant structures). <...> indicates average over all nodes (i.e., aperture patches).

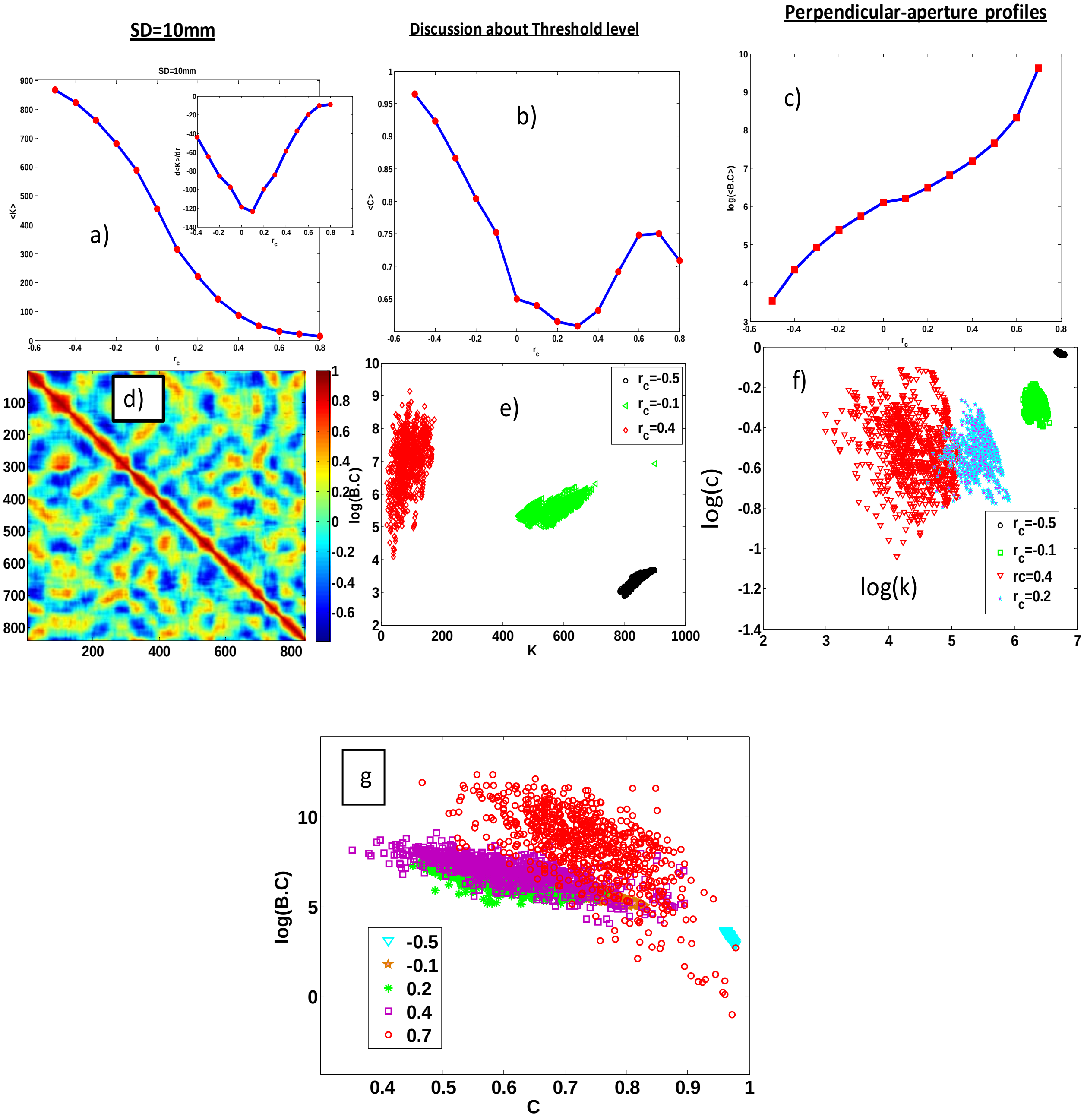

Figure S8. Sensitivity of the results with variation of threshold level at SD=10 mm :a) mean node's degree versus threshold level ( $r_c$ ) –the inset shows $\frac{dk}{dr_c}$ ;b)variation of <c> -mean clustering coefficient- against $r_c$ ;(c)variation of betweenness centrality(B.C) with $r_c$ ;d) correlation patterns over perpendicular profiles ;e) the variation of profiles in *B.C-k* space with threshold value ;f) the variation of profiles in *c-k* space with the threshold values; g) the variation of profiles in *B.C-C* space with threshold value.

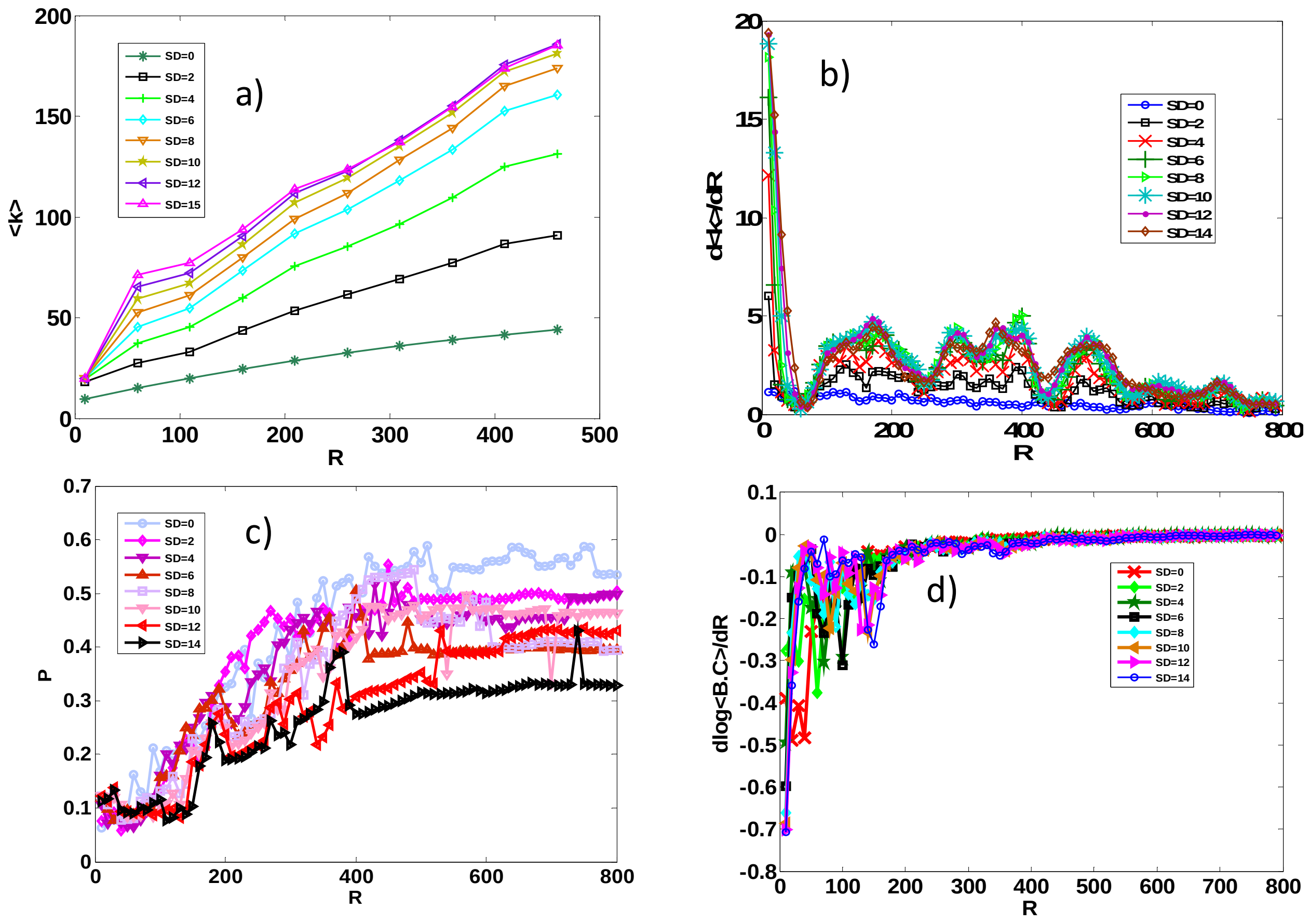

Figure S9. a) Mean degree of networks versus increasing *R* and for different shear slips; b) the variation of $\dfrac{d<k>}{dR}$ with R ;c) mean value of participation coefficient (P) of networks with variation of *R* and in successive slips ;d) $\dfrac{d\log<BC>}{dR}$ with variation of R and in successive slips.

To realization of geometrical friction networks, we need to choose a proper influence radius or effective zone. Sensitivity of degree density (average of edges), participation coefficient (P) and B.C with respect to variation of influence radius has been shown in Figure 9. The sensitivity of edges growth to R in last slips is higher than the initial displacements. Also rapid drop of $\dfrac{d<k>}{dR}$ in R<50, shows the fast variation (loosing) of structures within this zone. However, cooperation of links (P) in other modules shows a phase transition around R=150. Above this range the size of extra links to other modules rapidly increases. At higher influence radius it stabilizes until the extra-modules links spans to 50-60% of the maximum value. The same interval is observed in $\frac{d\log < BC >}{dR}$ while after R>150 the local information flow (B.C) is stabilized (Figure 9d). Then to obtain the maximum information from geometrical friction networks, it is better to choose the threshold influence value in the range out of high fluctuation.

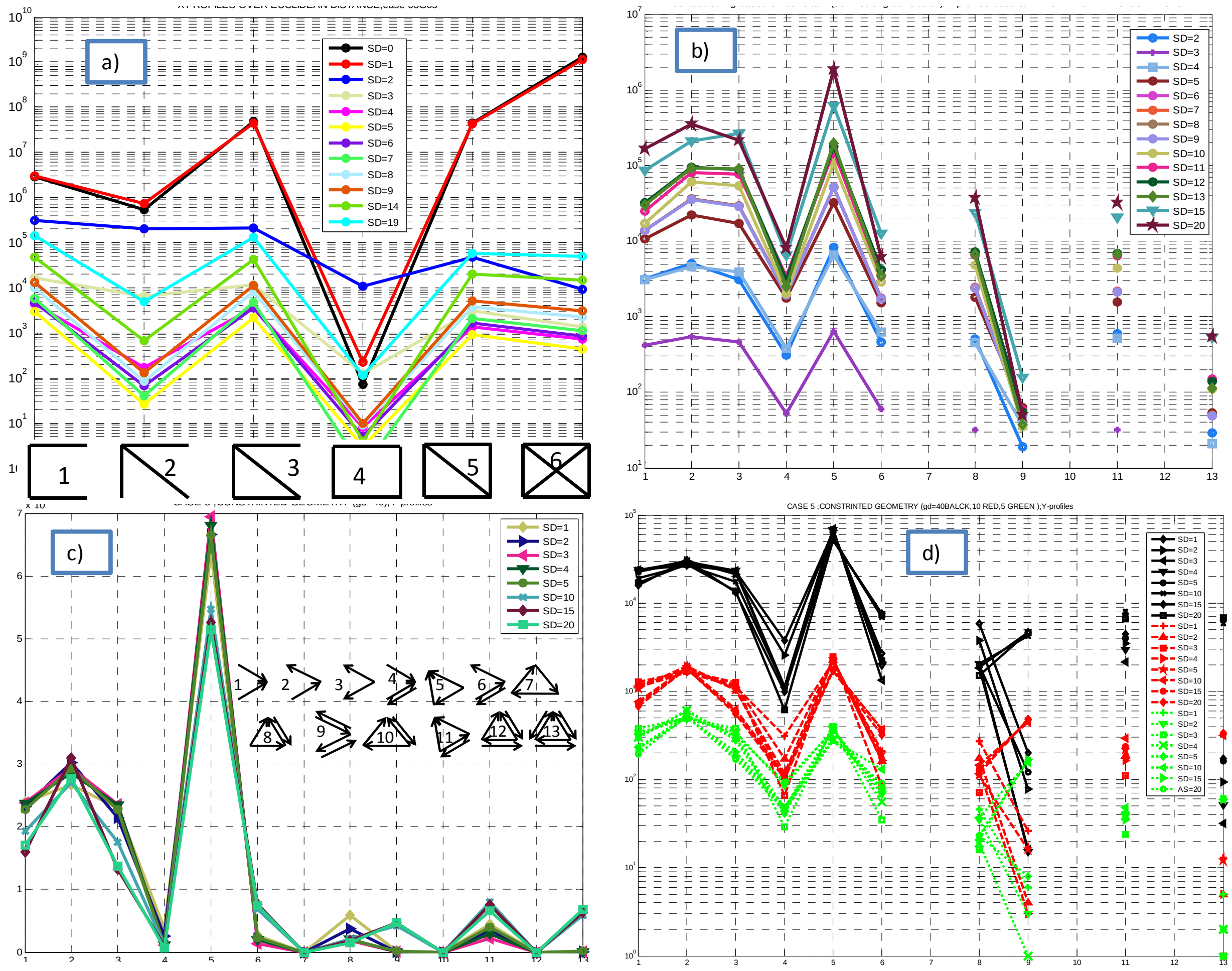

Figure S10 a) Distribution of 4-points sub-graphs with pre-set Euclidean measure over case 3 (perpendicular aperture profiles) ; b) Distribution of 3-points sub-graphs with pre-set correlation measure over case 3 (contact strings); c) Distribution of 3-points sub-graphs over case 3 with constrained geometrical distance and d) Comparison of sub-graphs ranks with different interactions' radiuses in geometrical distance over case 3(black: R=40 ;red: R=10 and green :R=5).

It is noteworthy; selection of the influence radius for enough large values does not change the patterns of motifs distribution (figure 10 d). However, smaller value of influence radius affects the motifs ranks, for example the peak point in motif with index5 is replaced with motif with index 2 (in directed friction networks). It is obvious; the directed friction networks show a self-similar pattern which seems is compatible with within module degree (Z)-X space (figure 6).

## Friction networks & Real-time contacts of frictional interfaces

We start with 2D interfaces, monitored at discrete time steps [0, 0.4, 0.75, 1, 1.2, and 1.4] *ms* ([7]-also see figure 11a and b). Transferring X-patches (then perpendicular to shear direction) to networks and plotting clustering coefficient revealed three distinct patterns of rupture evolution (Fig11.b) which is comparable with the previous results [7] . The three patterns correspond with sub-Rayleigh (1 and 3 in Fig12.b) and slow rupture (2)- figure 11b. In other words, movement of rupture tip is followed by fast variation of clustering coefficient. Furthermore, as well as aperture friction networks  and Eq.5, with considering 3 points cycles (*T*-triangle loops) versus node's degree from 0.4-1.4 ms show a power low scaling (Fig12.c):

$$T(k) \sim k^{\beta}, \tag{6}$$

where the best fit for collapsed data set reads $\beta \approx 2.1 \pm 0.4$ (we call coupling coefficient of local and global structures). Similar scaling law is obtained for other types of sub-graphs. Peculiarly, Euclidean distance yields same scaling law. With some mathematical analysis, one can show adding $m$ edges increase the number of loops with $\beta^2 m^{\beta}$ which indicates a very congested structure of global and local subgraphs during shear rupture. Also, we notice $C(k) \sim 2k^{\beta-2}$ so that for $\beta < 2$, a hierarchical structure is predicted [7].Next, we plot sub-graphs (*i.e.,* motifs) distributions (Fig.12d) which indicate a super-family phenomenon. A similar trend in all the rupture speeds indicates universality in energy flow in frictional interfaces which is characterized by friction laws. As It is clear we obtained the same trend in sub-graphs distribution as well as motifs ranks in aperture-friction networks. Remarkably, transition from sub-Rayleigh to slow rupture is correlated with a distinct jump in all types of the 4-points sub-graphs. With considering the variations of $<k>$ at the monitored interval ($10 \le k \le 30$) and abrupt change in loops in order of one magnitude, we conclude fluctuation of coupling coefficient induces remarkable growth of sub-growth. In figure 13and 14, we have illustrated our results over different sub-graphs, represents a similar scaling power law. Following the results, declares the universality of scaling law as it has been mentioned in Eq.5 and Eq.6. We can extract friction parameters as well as slip weakening distance or shear rupture energy in terms of friction networks parameters. Additionally, we can develop friction laws in terms of local and global patterns of friction networks as well as estimation of state variable in rate and rate friction laws.

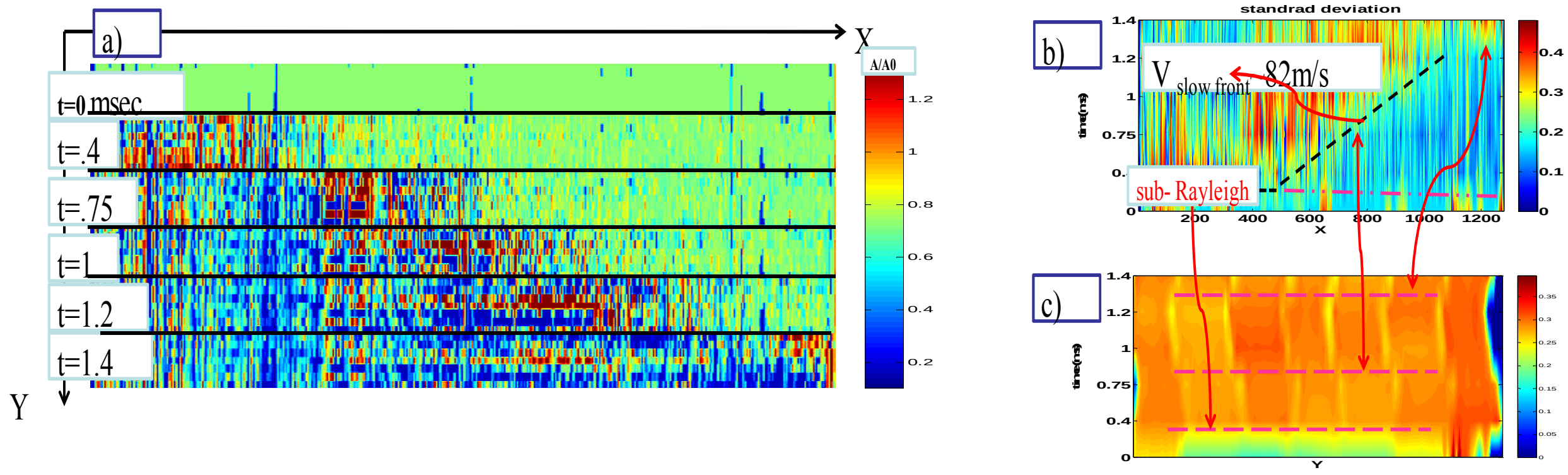

Figure S11. a) Evolution of real-time contacts through 6 time-windows in X-Y space of an interface. Each pixel corresponds with net contact area (relative)-Reconstructed figure from [7]. b) Standard deviations of contacts in X and c) Y directions.

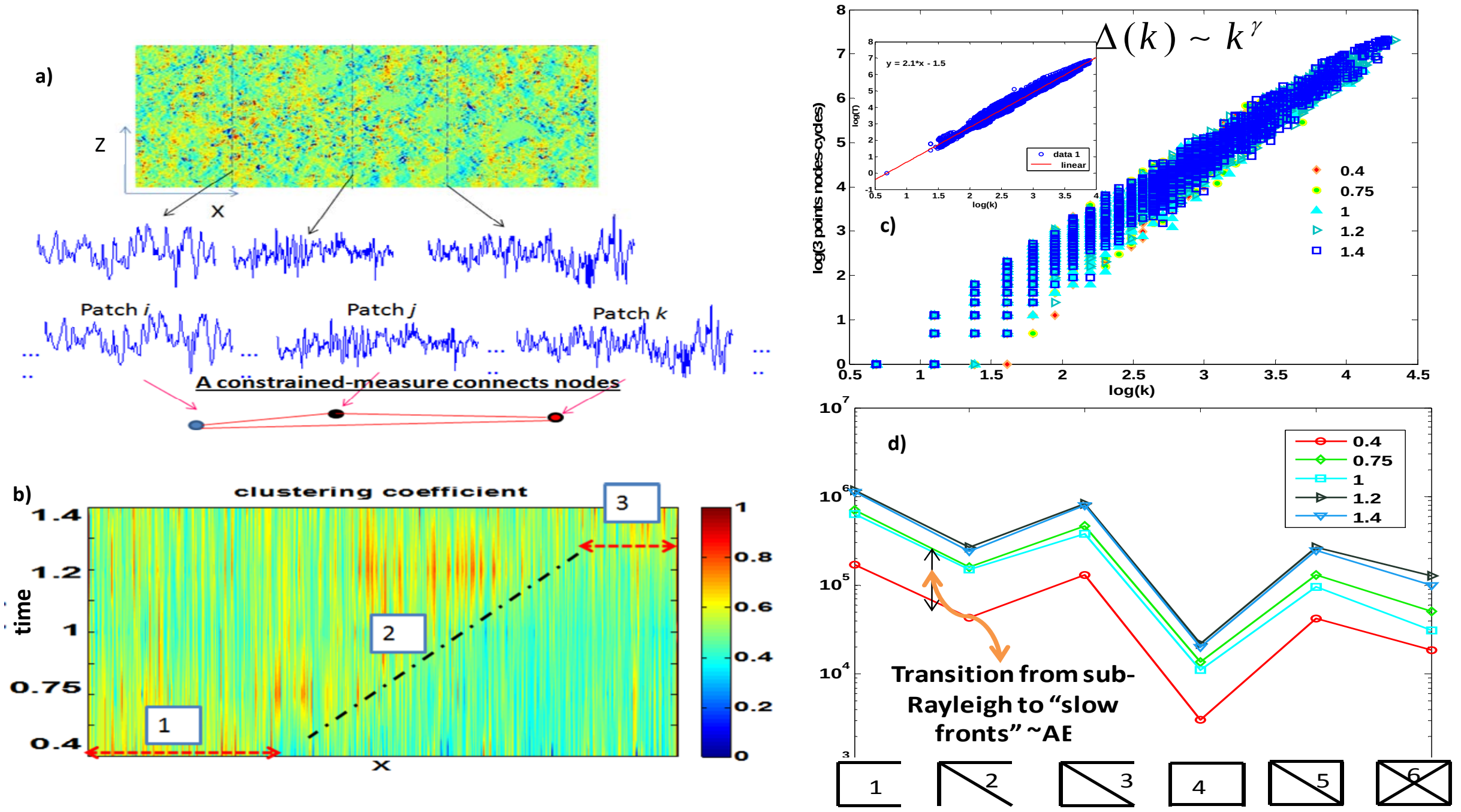

Figure S12. (a) A typical example of transferring contact areas into networks ,(b) mapping the dynamic relative contact areas in 2D [7] to networks space an plotting clustering coefficient as a fraction of triangles reveals the relatively precise rupture speed :three distinctive rupture speeds is compatible with mean contact area ;1 , 3 corresponds with sub-Rayleigh rupture and 2 is slow rupture ; (c) scaling of triangles in obtained networks with number of similar profiles (node's degree) ,expressed with power law relation with universal exponent (~2) ;(d) distribution of motifs of networks over different time steps shows a non-uniform /universal distribution ;transition from rupture(1) to rupture (2) occurs suddenly which is related to regular acoustic emission (high frequency waveform). http://eposters.agu.org/files/2011/12/2011-AGU-poster-Compatibility-Mode.pdf

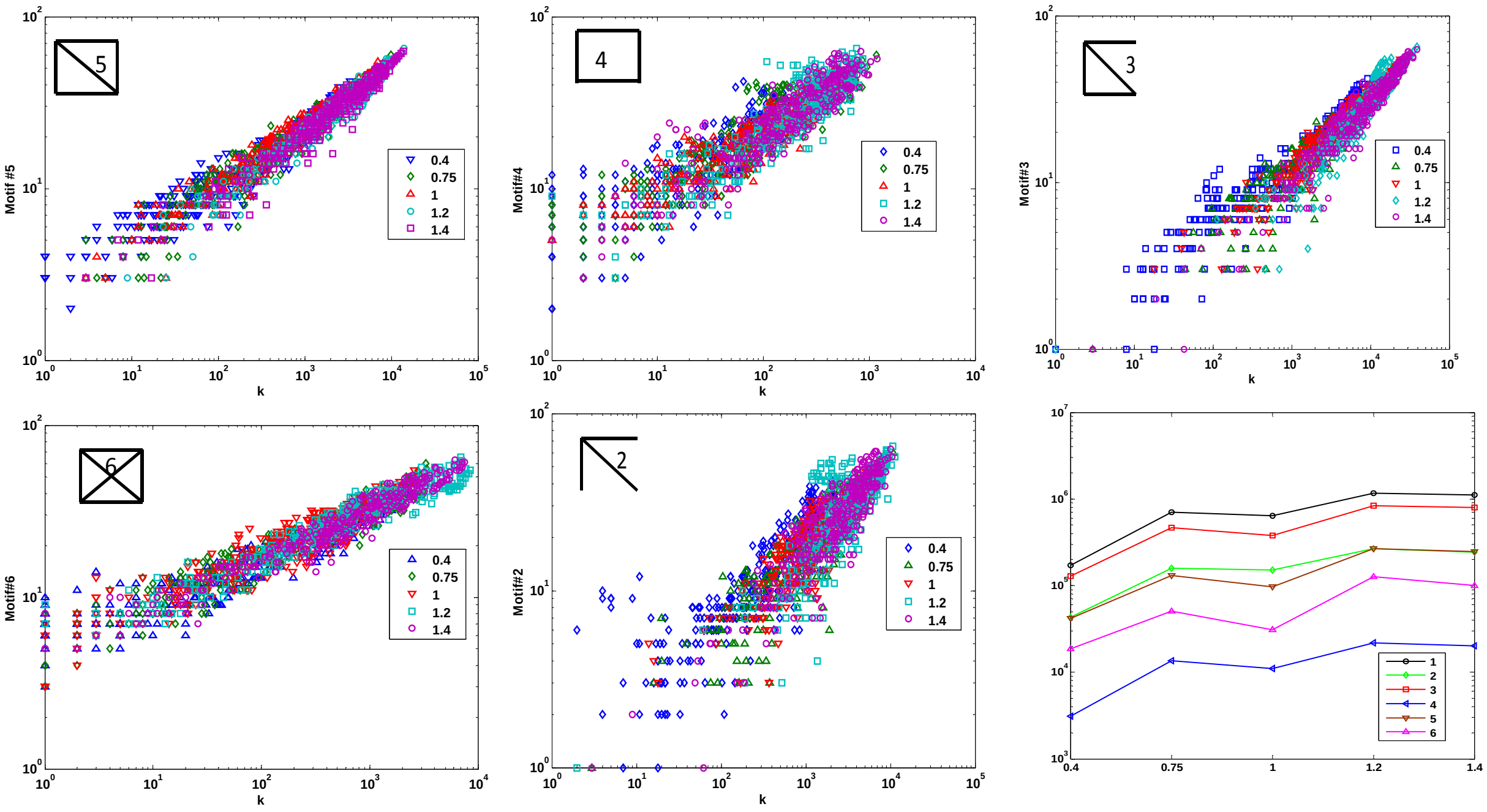

Figure S13.scaling law of motifs and node's degree in real-time contacts in [0.4, 0.75, 1, 1.2, 1.4] ms in stick time (before sliding) –data set from Prof. J.Fineberg (-The Racah Institute of Physics ,The Hebrew University of Jerusalem ,Givat Ram, Jerusalem , Israel).

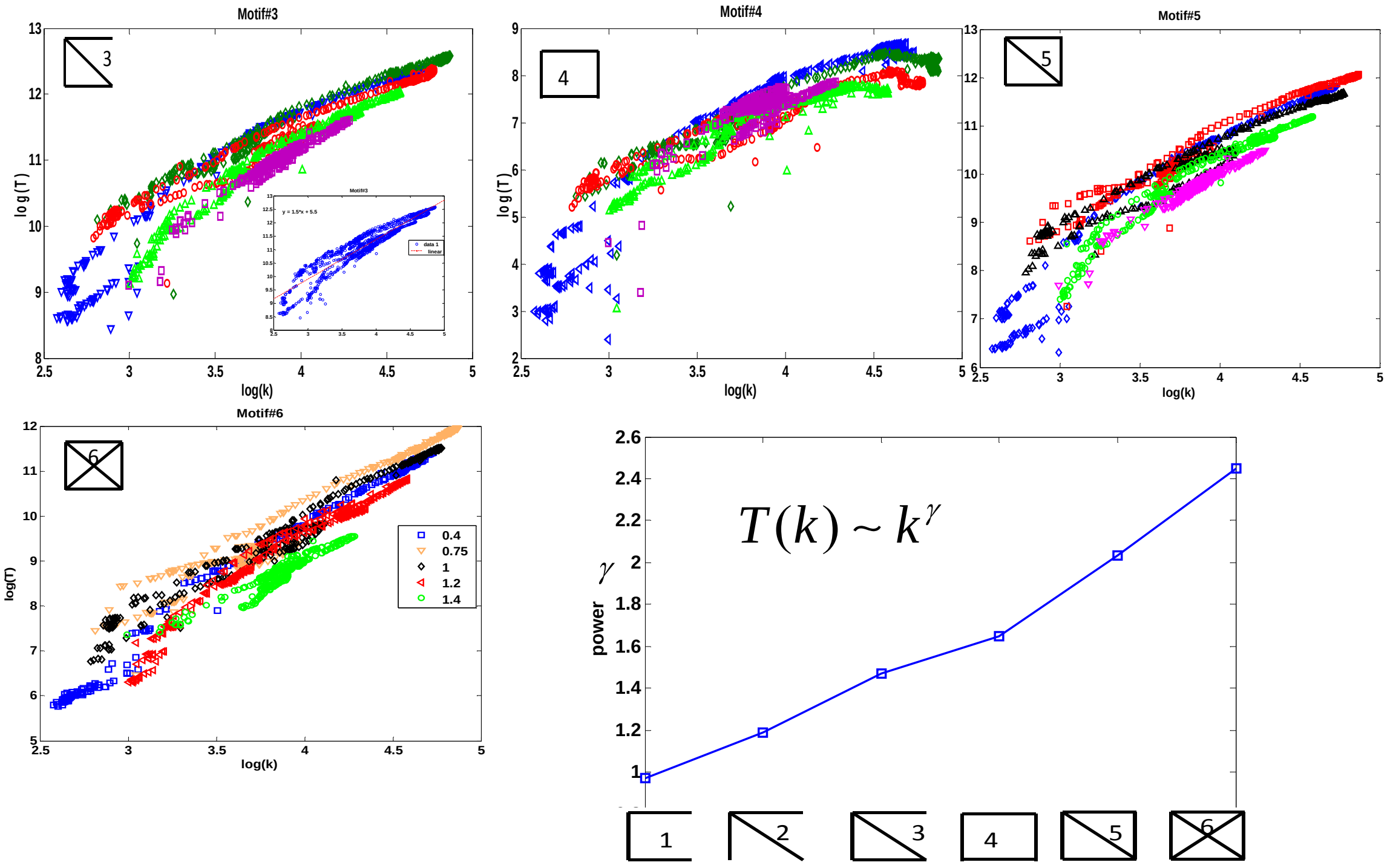

Figure S14. Scaling law when we are using Euclidean metric ; Increasing loops increases power of scaling laws.